\documentclass[aps,pre,showpacs,amsmath,amssymb,longbibliography, floatfix]{revtex4-2}
\usepackage{hyperref, subcaption, float, footnote}
\usepackage{amsmath, amssymb, latexsym, epsfig, graphics, epsf, import}
\usepackage[dvipsnames]{xcolor}

\usepackage{graphicx,float, hyperref, cleveref, import, subcaption, setspace}
\newcommand{\ea}[1]{\left\langle#1\right\rangle} 
\newcommand{\cA}{\mathcal{A}} 
\newcommand{\eq}[1]{Eq.~(\ref{#1})} 

\newcommand{\be}{\begin{equation}}
\newcommand{\ee}{\end{equation}}

\newcommand{\Figs}[1]{Figures~\ref{#1}}
\newcommand{\fig}[1]{Fig.~\ref{#1}}

\graphicspath{ {./figures/} }
\newcommand{\mycomment}[1]{}

\begin{document}

    \title{Property-dependent material times}
    \date{\today}
    \author{Aude Y. Amari}\email{aude.amari@gmail.com}
    \affiliation{\textit{Glass and Time}, IMFUFA, Department of Science and Environment, Roskilde University, P.O. Box 260, DK-4000 Roskilde, Denmark}
    \author{Lorenzo Costigliola}\email{lorenzo.costigliola@gmail.com}
    \affiliation{\textit{Glass and Time}, IMFUFA, Department of Science and Environment, Roskilde University, P.O. Box 260, DK-4000 Roskilde, Denmark}
    \author{Jeppe C. Dyre}\email{dyre@ruc.dk}
    \affiliation{\textit{Glass and Time}, IMFUFA, Department of Science and Environment, Roskilde University, P.O. Box 260, DK-4000 Roskilde, Denmark}

\begin{abstract}
    This paper analyzes simulations of the physical aging following sizable temperature up‑jumps from equilibrated slowly relaxing states. Jumps from the temperatures 0.43 and 0.37 to 0.48 are considered for a binary Lennard-Jones mixture. The Tool-Narayanaswamy (TN) concept of universal material time was recently shown to becomes less effective at rationalizing the aging response for such large jumps [Amari \textit{et al.}, Phys. Rev. E \textbf{113}, 045411 (2026)]. The current work investigates whether the effectiveness of the TN formalism is improved by parameterizing each observable by its ``own'' material time. This is done for the potential-energy time-autocorrelation function, the self part of the intermediate scattering function, and the time-dependent mean-square displacement. As part of the investigation a detailed analysis is carried out of how well the triangular relation is obeyed, a necessary condition for defining a material time. We conclude that for all properties the data collapse best as a function of a given property's own material time. The degree of improvement varies significantly among the three properties, however; it is most notable for the mean-square displacement. 
    \end{abstract}
\maketitle
\vspace{1cm}

\section{Introduction}\label{sec:introduction}

Physical aging refers to the gradual change of material properties over time due to rearrangements of its atomic or molecular constituents, which is qualitatively different from aging that involves chemical reactions. Physical aging is well described by the Tool-Narayanaswamy (TN) (or  Tool-Narayanaswamy-Moynihan (TNM)) formalism, which has been applied to materials spanning from metallic glasses \cite{rut17, son20} to oxide glasses \cite{nar71, scherer, mauro},  glasses of organic molecules \cite{ols98,hec10,roe19,rie22}, and polymers \cite{str78a,hod95, can13, mck17}. Aging is observed after an amorphous material close to its glass transition has been driven out of equilibrium by electrical \cite{Richert2023, Johari2013}, mechanical \cite{Lee2010, Bennin2020}, irradiation \cite{Sun2020, Alfinelli2023}, temperature \cite{Mehri2022, hec10} and/or pressure changes \cite{Di2004, Phan2020}. The study of physical aging is relevant both for modeling the manufacturing process and in the assessment of long-time properties of products in their later use. For this reason, the TN formalism has been commonly used in industry for decades \cite{scherer, mauro, Luthra2007, Lyubimova2016}.

Thermal treatments involved in production and during the lifetime of a product usually involve quite complex protocols. Physical aging is best studied scientifically 
in the conceptually simpler case of an instantaneous temperature jump, starting from an equilibrium state at some temperature and ending at a new temperature that is held constant while some property is being monitored as the system  relaxes toward the equilibrium (metastable) liquid state. This protocol is termed ``ideal'' if temperature is stabilized throughout the sample before any significant relaxation has taken place. Due to the generally slow heat conduction, an ideal temperature jump is challenging to realize in experiments, though it is possible to approach an ideal jump by having at least one very small sample dimension (reducing the heat-diffusion time) \cite{hec10, rie22, hen24}. In simulations, on the other hand, an  ideal jump is implemented easily by changing the thermostat temperature. In both experiments and simulations, temperature up and down jumps of same magnitude, $\Delta T$, to the same final temperature, $T_{0}$, result in quite different responses unless $\Delta T$ is very small. For a down jump starting at $T_{0} + \Delta T$, the energy decreases with time, causing a gradual slowing down of the relaxation toward equilibrium. In that case, aging is said to be "self-retarded" and it is stretched-exponential like. In contrast, an up jump starting at $T_{0} - \Delta T$ is "self-catalyzed" as relaxation events ease further relaxation, inducing a compressed-exponential behavior \cite{mck17, kov63, mal24, nis20}. Up and down jumps thus display the \textit{asymmetry of approach} symptomatic of non-linearity \cite{kov63,scherer,mck17}. This feature originates in the significant change in relaxation time manifested in the generally strong temperature dependence of the equilibrium relaxation time of glass-forming liquids \cite{dyr06,ber11}. Note that the time-dependent behavior of the system also implies that the time-translation invariance (TTI) characterizing equilibrium is broken.

\begin{figure}[ht]
    \centering
    \includegraphics[width=0.7\linewidth]{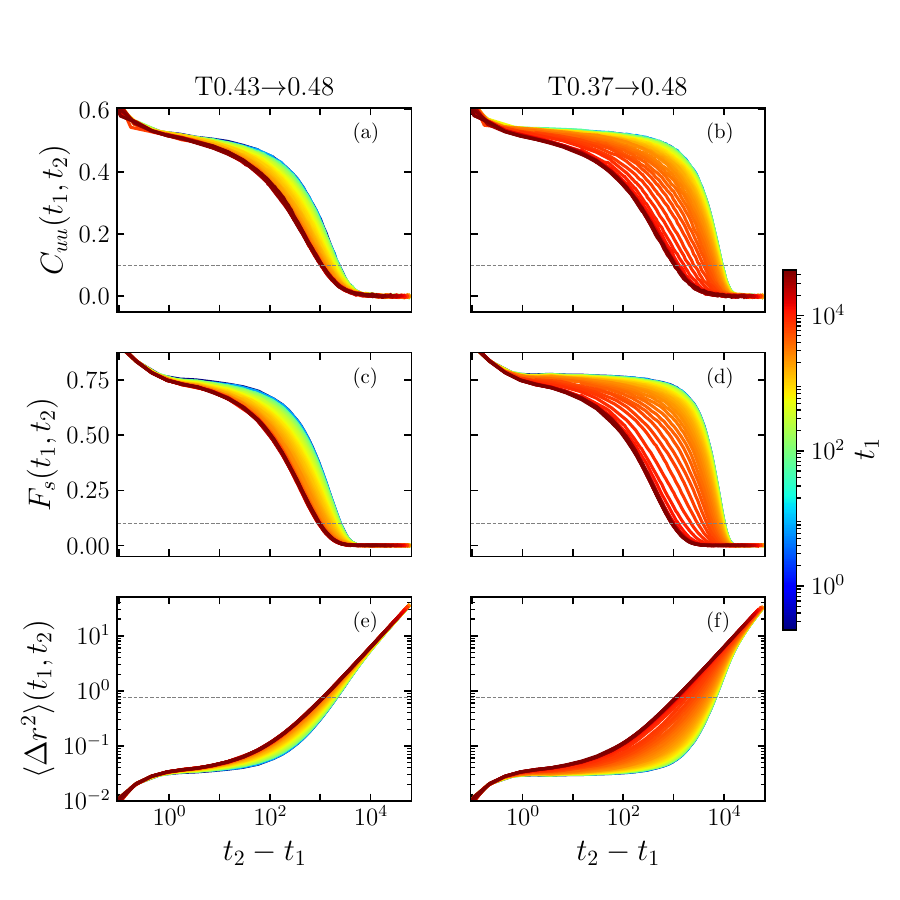}
    \caption{Relaxation following two temperature-up jumps to $T=0.48$ of a binary LJ mixture. The explicit dependence of the two-time functions on the waiting time $t_1$ of the two-time functions demonstrates that time-translation invariance is broken. 
    (a-b) Single-particle potential-energy time-autocorrelation function (normalized), 
    (c-d) incoherent intermediate scattering function (ISF), and 
    (e-f) mean-square displacement (MSD) during aging,
    all plotted as functions of the time difference $t_2 - t_1$ for different values of $t_1$ ($t_2\ge t_1$). 
    The curve colors vary from blue for small values of $t_1$ to red as this value increases. Data for a moderate and a large jump are shown in the left and right columns, respectively. The curves become identical at large $t_1$ (red) for the two jumps because they share the same final temperature. These figures present data obtained in Ref. \onlinecite{Amari2026} upon which this paper's analysis is based. }
    \label{fig:raw_data}
\end{figure}

Figure \ref{fig:raw_data} plots simulation results for three aging quantities upon two temperature up jumps applied to a binary Lennard-Jones (LJ) system, both ending at $T=0.48$ (in the expressions given below, angular brackets indicate ensemble averaging, which must be used instead of the usual starting-time averaging because TTI does not apply during aging). The three properties are:

\begin{itemize}
    \item The A particle potential-energy time-autocorrelation function $C_{uu}(t_1, t_2)$ defined by
    \be\label{eq:Cuu}
    C_{uu}(t_1,t_2) \equiv \frac{1}{N_A\,\sqrt{ \nu(t_1) \, \nu(t_2)}} 
    \left\langle\sum_{i=1}^{N_A} \Delta u_i(t_1) \Delta u_i(t_2)\right\rangle\,,
    \ee
    which measures how similar the potential energy of a particle at time $t_1$ is to itself at a later time $t_2$. Here $\Delta u_i(t) = u_i(t) - \ea{u(t)}$ in which  $u_i(t)$ is the potential energy of an A particle $i$ defined as $u_i = \frac{1}{2}\sum_{j=1}^{N} u_{ij}(r_{ij})$ where $u_{ij}(r_{ij})$ is the LJ pair potential, $u_{ij}(r_{ij})= 4 \epsilon_{ij}\Big[ \Big(\frac{\sigma_{ij}}{r_{ij}}\Big)^{12} - \Big(\frac{\sigma_{ij}}{r_{ij}}\Big)^{6} \Big]$, $\nu(t) \equiv \left\langle\sum_{i=1}^{N_A} (\Delta u_i(t))^2\right\rangle/N_A$, and $N_A$ is the number of A particles. We henceforth usually leave out the word ``normalized'' when speaking about time-autocorrelation functions and refer to $C_{uu}(t_1,t_2)$ simply as the potential-energy time-autocorrelation function. 

    \item The A-particle incoherent intermediate scattering function (ISF) $F_s(t_1, t_2)$ defined as 
    \be
    F_s(t_1, t_2) = \Bigg\langle \frac{1}{N_A} \sum_{i=1}^{N_A} \cos({\bf q}\cdot({\bf r}_i(t_2)-{\bf r}_i(t_1)) \Bigg\rangle\,,
    \ee
    which calculates the average Fourier transform of the displacements between times $t_1$ and $t_2$ at the wavenumber corresponding to the first peak on the A-particle radial distribution function. In practice, we averaged over three orthogonal wave vectors $\bf{q}$ in order to improve the statistics. Following the standard practice, we identified $\bf{q}$=7.25 from the first maximum of the A particle structure factor at the final temperature (this value remains virtually constant through the simulation). 

    \item The A-particle mean-square displacement (MSD) $\ea{\Delta r^{2}}(t_1, t_2)$ defined by
    \be
    \ea{\Delta r^{2}}(t_1, t_2) = \Bigg\langle \frac{1}{N_A}\sum_{i=1}^{N_A}({\bf r}_i(t_2) - {\bf r}_i(t_1))^{2} \Bigg\rangle\,.
    \ee
\end{itemize}
All quantities above and henceforth are given in A-particle LJ units. We do not retain in this study the two quantities considered in Ref. \onlinecite{Amari2026} with time-autocorrelation functions that are non-monotonic as function of $t_2$ for fixed $t_1$, the dynamical susceptibility and the non-Gaussian parameter. This is because in \eq{eq:def_clock} below, $\phi_{eq}^{\cA}$ is a single-valued function, and the non-monotonicity of those functions makes it impossible to translate them into a well-defined monotonic material-time difference $\xi_2-\xi_1$.

Both jumps presented in \Cref{fig:raw_data} go, as mentioned, to the same final temperature $T_{0}=0.48$. In the left column the initial temperature is $0.43$ and the final temperature is $T=0.48$, corresponding to a relaxation-time variation of roughly 10; in the right column the final temperature is the same but the initial temperature is $0.37$, corresponding to a  relaxation-time variation of roughly 10000. The observables are plotted as functions of the time difference $t_2-t_1$ parametrized by $t_1$, with curves going from bluish to reddish indicating increasing values of the ``waiting time'' $t_1$. The explicit $t_1$ dependence illustrates that TTI is broken. 

The TN model predicts that the non-linearity of aging can be entirely contained in a time reparameterization by introducing a material time, $\xi=\xi(t)$, such that all temperature jumps (also to different temperatures) result in the same normalized relaxation function as a function of $\xi$. As a consequence, for a general temperature protocol the relaxation response can be expressed as a linear convolution integral over the thermal history parametrized by $\xi$ \cite{nar71, scherer, hec15, dou22}. The success of the TN formalism in linearizing aging responses has been confined for small and moderate temperature jumps in both experiments \cite{Richert2018, Sellares2005} and simulations \cite{Mehri2022, Giovambattista2005}. For large jumps, the material time collapses the data less perfectly \cite{Amari2026, hen24, Moch2024}, however, although in some experiments TN formalism interestingly does work well in the ``single-parameter aging'' version even for quite large jumps \cite{hec24}. 

In Ref. \onlinecite{Amari2026}, two possibilities to improve the collapse of data for large temperature jumps were mentioned: 1) Defining a separate material time for each aging property; 2) Resorting to spatially resolved ``local'' clocks. The current paper investigates the first option, i.e., whether improved collapse is obtained for a given property as a function of material time by using ``its own'' material time. We start by recalling briefly in \Cref{sec:system} the system simulated in Ref. \cite{Amari2026} from which we analyze the data. Then in \Cref{sec:triang} we discuss the Cugliandolo-Kurchan triangular relation introduced in Ref. \cite{cugliandolo_1994} and its relation to the material-time concept, while \Cref{sec:check_time_invariants} defines different material times for each of three aging properties. Finally, in \Cref{sec:collapse} we examine how well the data collapse by implementing those reparametrizations.

\section{System simulated}\label{sec:system}
We use the data presented in Ref. \onlinecite{Amari2026} where a binary Lennard-Jones (mBLJ) model was simulated that has the same interaction parameters as the Kob-Andersen model \cite{kob95} but is less susceptible to crystallization. The mBLJ system thus consists of a mixture of 80\% large particles (A), and 20\% small particles (B) interacting via a LJ pair potential $v(r) = 4\epsilon \Big((\sigma_{\alpha\beta}/r)^{12} - (\sigma_{\alpha\beta}/r)^{6}\Big)$ with $\alpha, \beta \in \{A, B\}$, and $\sigma_{AA}=1, \; \sigma_{BB}=88, \; \sigma_{AB}=0.8, \; \epsilon_{AA}=1, \; \epsilon_{BB}=0.5, \; \text{and} \; \epsilon_{AB}=1.5$. Instead of the shifted-potential cutoff of the standard Kob-Andersen model, a shifted-force cutoff is used (located at 1.5 for the AA and BB interactions and at 2.5 for the AB interaction). 

Crystallization proceeds via A-particle phase separation \cite{ped18, Zanotto2020}, a process that is impeded by the shifted-force weakening of the attractions between identical particles. The result is a system that is at least 100 times less prone to crystallization \cite{schroder_solid-like_2020}, but which otherwise has virtually the same properties as the standard Kob-Andersen system. As mentioned, all considered properties are computed for A particles only as they dominate the relaxation dynamics. All simulations were carried out in the NVT ensemble using a Nosé-Hoover thermostat with relaxation time $\tau_{th}=0.2$, time step $dt=0.005$, each simulation lasting a total of $2^{23}$ time steps. The data presented are all averaged over 50 independent simulations.

\section{Triangular relation and material-time definition}\label{sec:triang}

Independently from the experimentally motivated TN formalism, Cugliandolo and Kurchan (CK) in 1994 developed a theory of aging in the context of spin-glasses \cite{cug94}, a theory that was later realized to be of much more general validity. The core idea is that although aging obviously breaks TTI, the relaxation dynamics presents a quasi-symmetry termed ``time-reparametrization softness'' (or approximate invariance). 

CK considered huge temperature down jumps, basically from infinite to zero temperature, and focused on describing the aging in terms of a time-reparametrization invariant action derived from a mean-field description. This is in stark contrast to the TN approach that was designed for describing continuous temperature variations and relatively small temperature jumps \cite{nar71,scherer}. Moreover, the TN focus was on linearizing the aging response, i.e., to reduce the description to one of standard linear-response theory with a translationally invariant convolution kernel when time is replaced by material time. This implies, in particular, that the normalized response to a temperature jump is a unique function of the material-time difference, i.e., independent of both sign and magnitude of the jump. Despite the significant differences between the TN and CK starting points, the two approaches are fully compatible, as we now demonstrate.

First, we remind the reader about the CK \textit{triangular relation} \cite{cug94}. Consider the time-autocorrelation function $C^\cA_{12}\equiv C^\cA(t_1, t_2)= \langle\cA(t_1)\cA(t_2)\rangle$ ($t_1<t_2$), of a quantity $\cA(t)$ with equilibrium average value equal to zero (otherwise one subtracts the equilibrium value). We assume that $C^\cA_{12}$ is positive and converges monotonously to zero for $t_2-t_1\to\infty$. Consider first the case of thermal equilibrium at some temperature. Here by TTI $C^\cA_{12}$ depends only on $t_2-t_1$ (which is of course not the case during aging). This implies that the time difference $t_2-t_1$ is uniquely determined by $C^\cA_{12}$. Likewise, $t_3-t_2$ is a function of $C^\cA_{23}$ and $t_3-t_1$ is a function of $C^\cA_{13}$. Since $t_3-t_1=(t_2-t_1)+(t_3-t_2)$, this means that in equilibrium $C^\cA_{13}$ is a function of $C^\cA_{12}$ and $C^\cA_{23}$. In other words:

\begin{equation}\label{eq:triang}
    C^\cA_{13} = F^\cA(C^\cA_{12},\, C^\cA_{23})\,.
\end{equation}
This is the triangular relation \cite{cug94, kur23}; note that the function $F^\cA$ in general depends on the property $\cA$. The idea is now that the very same relation applies during aging \cite{cug94,dyr15,dou22,kur23}. Not that while we here regard the triangular relation as a natural extension of a (trivial) equilibrium identity, the original CK argument did not refer to equilibrium at all, but instead justified \eq{eq:triang} from a mean-field description that separates short time scales from the long times relevant for aging \cite{cug94}.

The TN concept of a material time $\xi(t)$ connects to the triangular relation as follows. The TN formalism assumes that the normalized relaxation function following a temperature jump is determined by the material time that has passed since the jump was initiated. An obvious generalization of this is that any time-autocorrelation function $C^\cA_{12}$ likewise is a unique function of the material-time difference, $\xi(t_2)-\xi(t_1)$, providing the ``true'' measure of how long time has passed. Since the triangular relation refers to equilibrium at some reference temperature, in terms of the equilibrium time-autocorrelation function at $T_0$, $C^\cA_{eq}(t_2-t_1)$, we can now define a material time by

\be\label{eq:def_clock}
    C^\cA_{12} = \phi^\cA_{eq}(\xi^\cA(t_2)-\xi^\cA(t_1))\,.
\ee
This definition embodies the triangular relation because -- in analogy with the above argument -- if $C^\cA_{12}$ is a function of the material-time difference $\xi(t_2)-\xi(t_1)$ and $C^\cA_{23}$ is a function of the material-time difference $\xi(t_3)-\xi(t_2)$, then the triangular relation implies that $C^\cA_{13}$ is a function of the material-time difference $\xi(t_3)-\xi(t_1)=(\xi(t_3)-\xi(t_2))+(\xi(t_2)-\xi(t_1))$. In other words, the triangular relation ensures consistency of the assumption that the time-autocorrelation function is determined by the material-time difference. Conversely, if the time-autocorrelation function is determined solely by the material-time difference, then the triangular relation must apply because knowledge of $C^\cA_{12}$ and $C^\cA_{23}$ implies knowledge of $\xi(t_2)-\xi(t_1)$ and $\xi(t_3)-\xi(t_2)$, which implies $\xi(t_3)-\xi(t_1)$ that in turn determines $C^\cA_{13}$.

Despite the concept of material time being straightforward, calculating this function from experiments is a very demanding task, and Narayanaswamy and subsequent authors obtained the material time only implicitly from the time evolution of the so-called fictive temperature \cite{NARAYANASWAMY_1971, Moynihan1976}. Recently, however, the material time has been directly extracted from experimental time-autocorrelation functions \cite{boh24}.

\section{Defining materials times}\label{sec:check_time_invariants}

We start by examining to which degree the triangular relation \eq{eq:triang} holds for the potential-energy time-autocorrelation function, the incoherent (self part) of the ISF, and the MSD.  \Cref{eq:triang} is verified for each of the three two-time autocorrelation function in \fig{fig:binning_algo} of the Appendix (in which details are given for how a distribution of $C^\cA_{13}$ values is obtained for ``pixel'' intervals of side lengths $\Delta C^\cA_{12}$ and $\Delta C^\cA_{23}$).

\begin{figure}
    \centering
    \includegraphics[width=\textwidth]{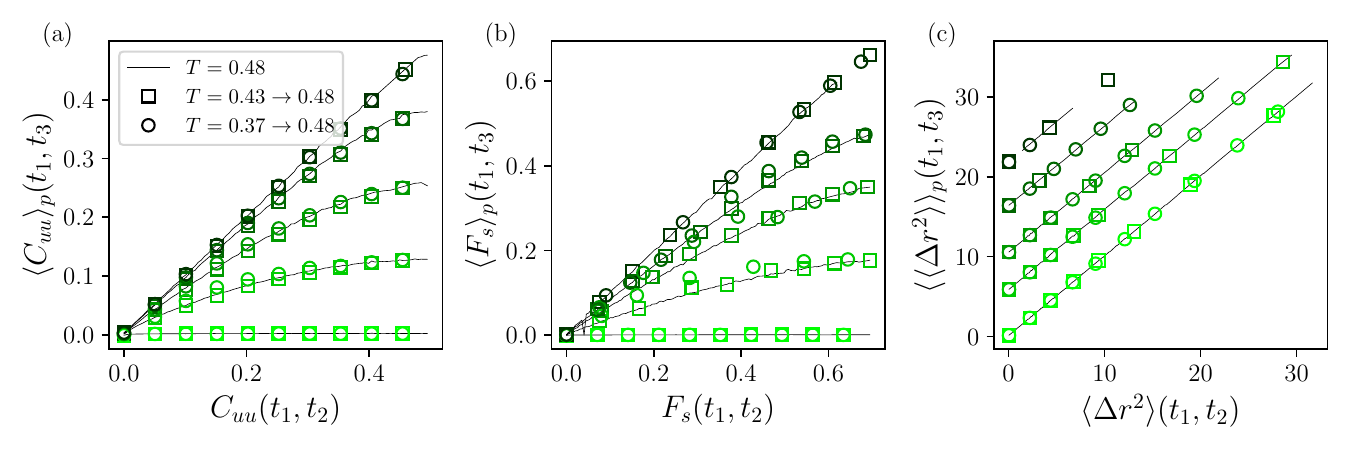}
    \includegraphics[width=\textwidth]{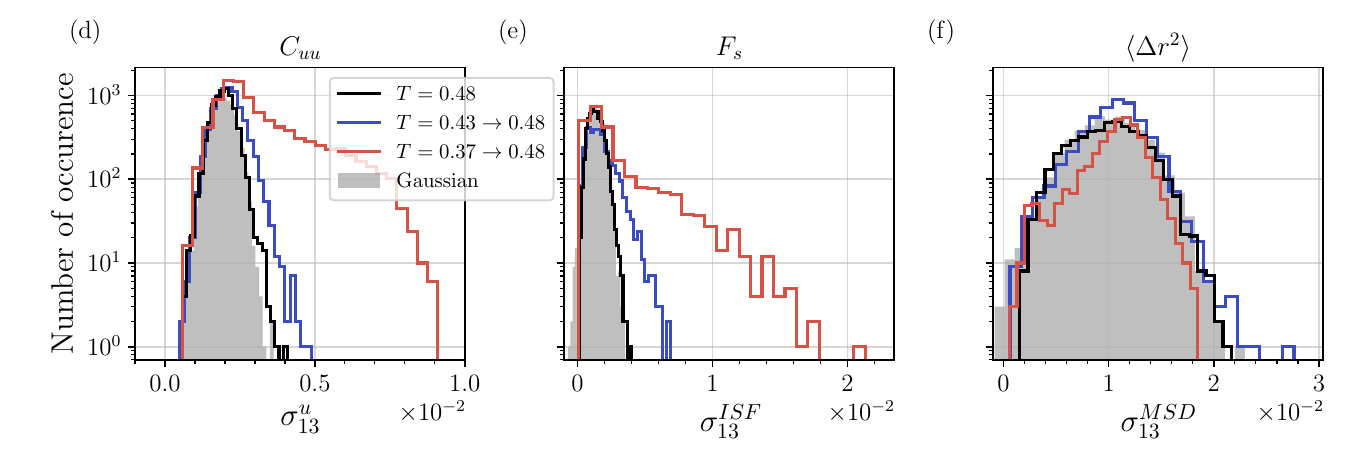}
    \includegraphics[width=\textwidth]{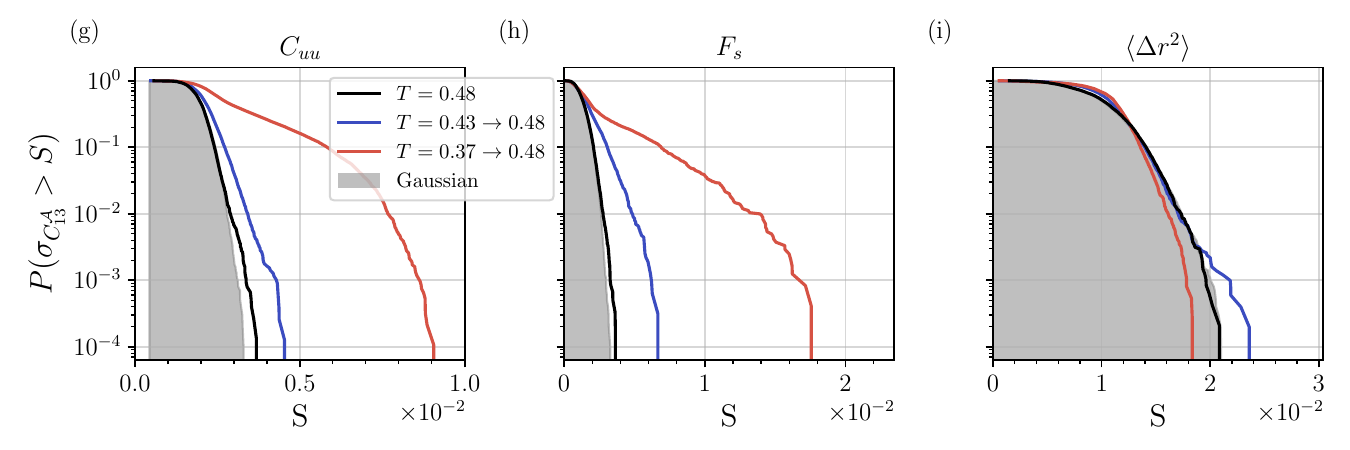}
    \caption{Triangular relation \eq{eq:triang} at equilibrium, as well as for a moderate and a large temperature up jump for: (a,d,g) the potential autocorrelation function, 
    (d,e,h) the self part of the ISF, and 
    (c,f,i) the MSD.
    (a-c) Per pixel average of $\ea{C^\cA_{13}}_p$ as a function of $C^\cA_{12}$ for fixed values of $C^\cA_{23}$ (light to dark green for small to large values of $C^\cA_{23}$). The shape of the relation $F^\cA$ in \eq{eq:triang} between the two-time functions at times $t_1, t_2$ and $t_3$ is identical at equilibrium and during aging but depends on the observable $\cA$.
    (d-f) Distribution of standard deviation values.   
    (g-i) Complementary Cumulative Distribution Function (CCDF) $P(\sigma_p>S)$ whose standard deviation $\sigma_{C^{\cA}_{13}}$ exceeds the value $S$, normalized by the total number of pixels, $n_{\sigma>0}$, plotted as a function of $S$. 
    (d-e,g-h) Deviations from \eq{eq:triang} are visible during aging as shown by the data deviating from the equilibrium baseline. 
    (f-i) For the MSD, \Cref{eq:triang} holds perfectly even for the largest jump.}
    \label{fig:triang_results}
\end{figure}

The triangular relation is investigated by binning a large number of values of $C^{\cA}_{13}$ in small pixels for given values of $C^{\cA}_{12}$ and $C^{\cA}_{23}$. The statistics of $C^{\cA}_{13}$ per pixel is presented in \fig{fig:triang_results}. In equilibrium \eq{eq:triang} is strictly satisfied and therefore the numerical noise from equilibrium can be used as a baseline for the data during aging. \Figs{fig:triang_results}(a-c) show the mean value per pixel of $C^\cA_{13}$ as a function of $C^\cA_{12}$ for fixed values of $C^\cA_{23}$. This is done for $C_{uu}$, ISF, and MSD at equilibrium (full line) and during aging after the moderate (open squares) and the large (open circles) temperature up jumps. The curves go from light to dark with increasing values of $C^\cA_{23}$. The orange curve in \fig{fig:binning_algo}(b) illustrates where one curve comes from. While the function $F^\cA$ depends on the observable $\cA$, it is identical at equilibrium and during aging as shown by the good collapse. 

Small fluctuations are visible in these plots between equilibrium and aging, but the general conclusion is that panels (a)-(c) are consistent with the triangular relation. A detailed study of deviations from the triangular relation is presented in Figs. \ref{fig:triang_results}(d-f), which show the distribution standard deviations values within a pixel $\sigma_{C^{\cA}_{13}}$. The quantity $\sigma_{C^{\cA}_{13}}$ is illustrated in the circular inset of \fig{fig:binning_algo}(b) of the Appendix. \Figs{fig:triang_results}(g-i) show the Complementary Cumulative Distribution Function (CCDF) of $\sigma_p$. These plots trace the probability $P(\sigma_{C^{\cA}_{13}}>S)$ that $\sigma_{C^{\cA}_{13}}$ is greater then a threshold value $S$. Both the histogram and the CCDF show the Gaussian distribution found at equilibrium, and the persisting deviations during aging for $C_{uu}$ and $F_s$, especially for the largest jump (red circles) for which $\max(\sigma_{C^{\cA}_{13}})$ reaches up to 2.3 and 5.6 times the maximum noise. For the MSD on the other hand, the deviations during aging remain close to the equilibrium baseline, meaning that \eq{eq:triang} holds perfectly. What is to be regarded as a ``large'' deviation from the triangular relation is debatable; thus one cannot easily say that \eq{eq:triang} does not hold for $C_{uu}$ and $F_s$ after the jumps from $T=0.37$. We conclude that the triangular relation is obeyed to a good approximation for all three two-time functions, implying that each of them may be used to define a material time. 

\begin{figure}[ht]
    \centering
    \includegraphics[width=0.85\linewidth]{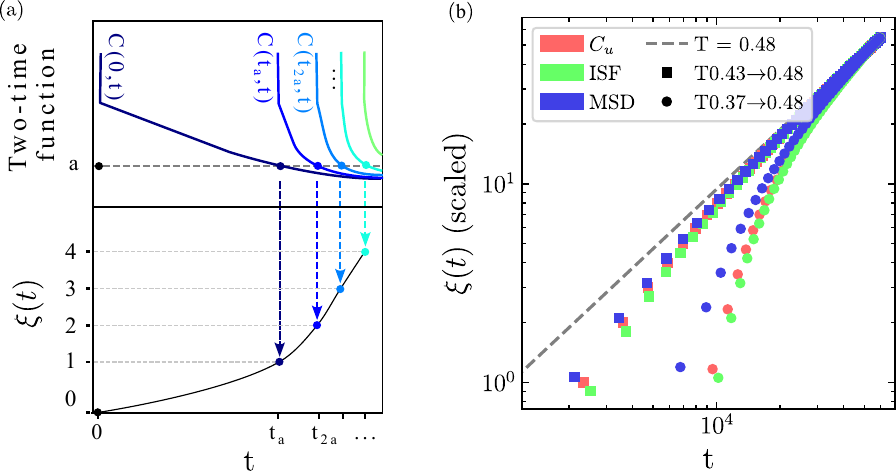}
    \caption{(a) Illustration of how the material times are defined and determined numerically. (b) Material time for the potential-energy time-autocorrelation function, incoherent ISF, and MSD, as functions of time for the two jump to $T=0.48$ starting from $T=0.43$ (circles) and $T=0.37$ (squares). The difference between the material times is especially visible for the large jump.}
    \label{fig:def_MT}
\end{figure}

For each of the three properties under study, a material time is next determined as follows. First a value $a$ is chosen; when the system has decorrelated to $a$, by definition one unit of material time has elapsed, i.e., $\xi^\cA(t_2)-\xi^\cA(t_1)=1$ when $C^\cA(t_1, t_2)=a$. The temperature jump is initiated at $t=0$, and we also initiate the material time by the condition $\xi^\cA(t=0)=0$. After identifying the time $t_a$ from $C^\cA(0, t_a)=a$, the next time $t_{2a}$ characterized by $\xi^\cA(t_{2a})=2$ is found from $C^\cA(t_a, t_{2a})=a$, i.e., $t_{2a}$ is the time it took for the second unit of material time to elapse. In this way $\xi^\cA(t)$ is constructed iteratively from each of the three two-point functions. $\xi^{\cA}(t)$ is defined up to a constant multiplicative factor; when \eq{eq:triang} holds perfectly, the choice of $a$ is arbitrary (as long as it is far enough from the short-time plateau). In practice, the collapse depends slightly on the choice of $a$ because \eq{eq:triang} is not perfectly satisfied (see Sec. C of the Appendix). We used $a=0.1$ for both $\xi^u$ and $\xi^{ISF}$ and $a=0.75$ for $\xi^{MSD}$; these choices ensure that one unit of material time corresponds to approximately $10^3$ time units at $T=0.48$ for all properties. Though this choice does not result in the very best data collapse, it allows a direct comparison across the three material times at the same stage of the system's relaxation. The horizontal dashed lines in \fig{fig:raw_data} indicate the selected values of $a$. Note that the current choice of $a$ to define the material time from the potential-energy time-autocorrelation function differs from that of Ref. \onlinecite{Amari2026} in which $a=0.4$. -- \Cref{fig:def_MT}(b) shows the three material times for the two jumps as functions of time in a log-log plot.  The curves are scaled slightly to agree at long time. 

\section{How well do the three material times collapse data?}\label{sec:collapse}

We have shown that the triangular relation is obeyed well enough for the potential-energy time-autocorrelation function, the incoherent ISF, and the MSD, that a material time can be defined from each of them. We carry on by considering the data presented in \fig{fig:raw_data} after reparametrization by these three material times.

\begin{figure}[ht]
    \centering
    \includegraphics[width=0.7\linewidth]{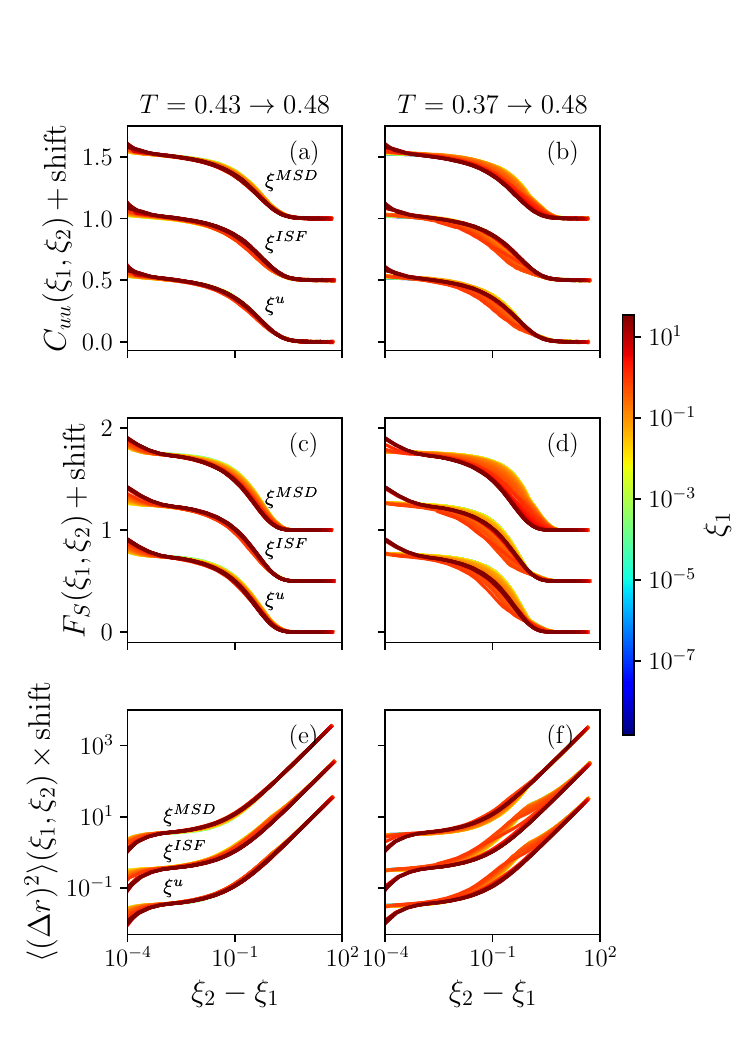}
    \caption{Two-time observables during aging after reparametrization by the material times defined from the potential-energy time-autocorrelation function $\xi^u$, $\xi^{ISF}$ from the ISF, and $\xi^{MSD}$ from the MSD, plotted as functions of the corresponding material time differences, $\xi(t_2)-\xi(t_1)$. The color of the curves vary from red for small values of $\xi(t_1)$ to blue as this value increases. 
    (a-b) Potential-energy time-autocorrelation function, 
    (c-d) self part of the ISF, 
    (e-f) MSD. Results for the moderate and large jumps are shown on the left column and right column, respectively. The data collapse is better for the moderate jump than for the large one, to a varying degree depending on the reparametrization.}
    \label{fig:3collapse_3MT}
\end{figure}

\Cref{fig:3collapse_3MT} investigates how the material time $\xi^\cA_i=\xi^\cA(t_i)$ defined from $\cA=C_{uu}$, ISF, MSD, collapse the two-time functions to a master curve when plotted as a function of $\xi^\cA_2-\xi^\cA_1$. As before, the curves vary from bluish to reddish as the value of the waiting time increases. At small $\xi^\cA_2-\xi^\cA_1$ values there is no collapse (as expected), while the collapse is better at later times, although to a varying degree. Not surprisingly, given that there must be perfect collapse in equilibrium for which $\xi^\cA(t)\propto t$ for all three properties, the moderate temperature jump systematically results in a better collapse than the large one. Interestingly, the bluish curves are not always the ones that deviate most from a collapse, an effect that is emphasized by our choice of $a$ far from the relaxation plateau (see Sec. C of the Appendix). 

\begin{figure}[H]
    \centering
    \includegraphics[width=0.4\linewidth]{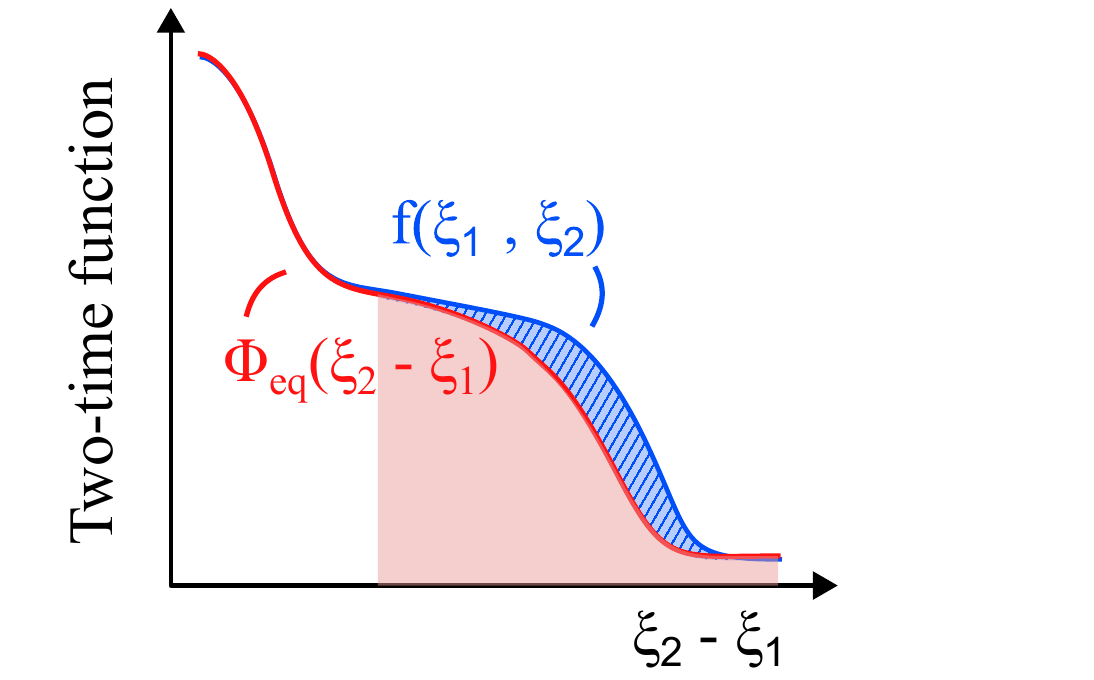}
    \caption{Schematic illustration of our measure of the collapse of aging data after the reparametrization of time defined by \eq{eq:def_L2norm}.}
    \label{fig:L2_visual}
\end{figure}

We proceed to introduce a quantitative measure of the degree of collapse. This is done by defining a measure termed $\text{col}[f](\xi_1)$ of the collapse of a function $f(\xi_1, \xi_2)$ at fixed $\xi_1$ onto the equilibrium two-time function $\phi_{eq}(\xi_2-\xi_1) = f(\xi_1\rightarrow\infty, \xi_2)$ reached at long waiting time. The collapse measure is defined by

\be \label{eq:def_L2norm}
    \text{col}[f](\xi_1) = \frac{ \| f  - \phi_\text{eq} \|_2 ( \xi_1) }{\| \phi_\text{eq}\|_2 (\xi_1)}
\ee
in which $\|f\|_2(x)=\Big[\int_0^{\infty} (f(x,y))^2 dy \Big]^{1/2}$ is the Euclidean or $l^2-$norm of the mapping $y\to f(x,y)$ for fixed $x$. The numerator quantifies the deviation between $f(\xi_1, \xi_2)$ and $\phi_{eq}(\xi_2-\xi_1)$ (blue area in \fig{fig:L2_visual}), which is normalized by the $l^2-$norm of the final equilibrium curve (red area). The smaller $\text{col}[f](\xi_1)$ is, the better is the collapse of the curve at $\xi_1$ as shown in \fig{fig:L2_visual}. 

\begin{figure}[H]
    \centering
    \includegraphics[width=0.7\linewidth]{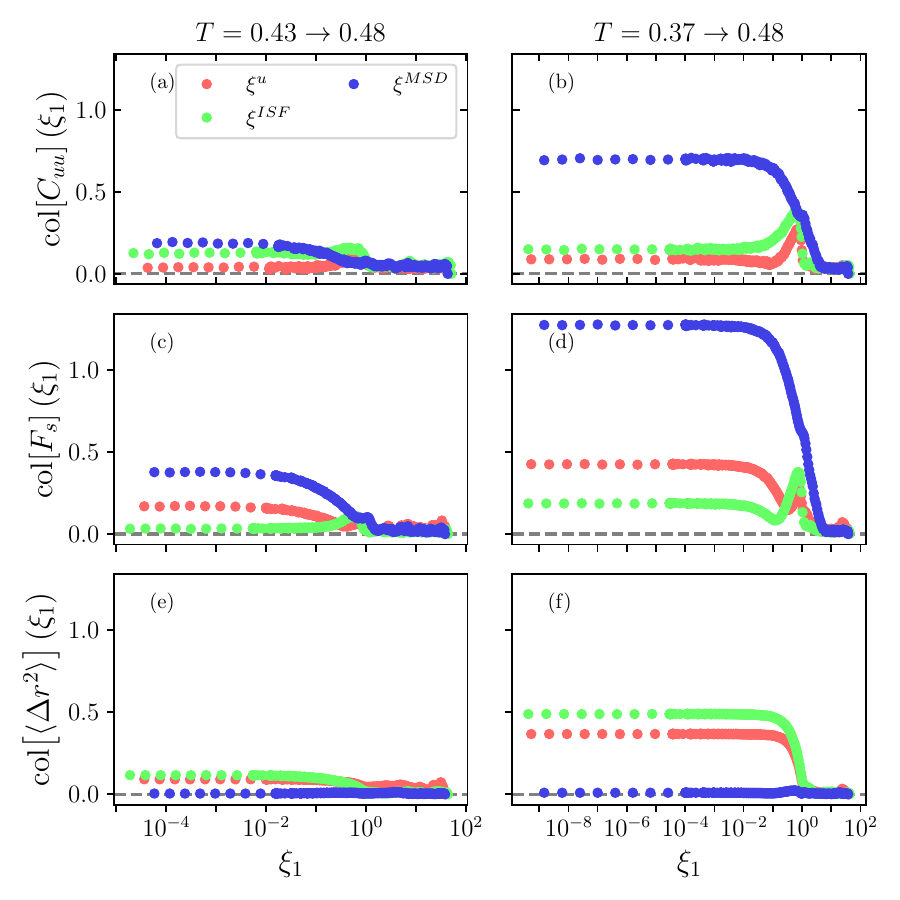}
    \caption{Measure of the deviation from collapse in linear scale as a function of $\xi_1$ for: 
    (a-b) the potential-energy time-autocorrelation function, 
    (c-d) the self part of the ISF, and 
    (e-f) the MSD, 
    have been reparameterized by the three material times. The best collapse is always obtained by reparametrizing a two-time function by its own material time, as shown by the smallest values of $\text{col}[f](\xi^\cA)$ obtained from $\xi^\mathcal{B}$ when $\mathcal{A}=\mathcal{B}$, but the improvement is not always significant.}
    \label{fig:main_summary}
\end{figure}

\Cref{fig:main_summary}(a) shows one of the main result of this paper by plotting $\text{col}[f](\xi_1)$ as a function of $\xi_1$. The left and right columns show the moderate and large jumps, respectively; from top to bottom is investigated how well $C_{uu}$, ISF, and MSD collapse after reparametrization by $\xi^u$, $\xi^{ISF}$, and $\xi^{MSD}$ in red, green and blue, respectively. 

In all three cases, $\text{col}[f](\xi_1)$ as expected becomes small at long $\xi_1$, i.e., as the system approaches equilibrium. The collapse is generally better for the moderate jump; it tends to degrade for the large jump although this happens to a varying degrees as previously noted. In all three cases the best collapse is obtained for a given property after reparametrizing by its own material time, although it should be emphasized that the improvement is not always sizable. Thus on the one hand the MSD is poorly collapsed by $\xi^u$ and $\xi^{ISF}$, and the improvement is significant if it's own material time is used instead. On the other hand, while $C_{uu}$ is also better collapsed by its own material time, the improvement compared to using $\xi^{ISF}$ is here not very significant. Curiously, the collapse obtained with a property's own material time is occasionally worse than with other reparameterization for some values of $\xi_1$, as can be seen in \fig{fig:main_summary}(d). This effect changes somewhat with the choice of $a$ because the triangular relation is not strictly satisfied; it generally happens before a unit of material time has elapsed. We cannot explain, however, why that is.  


\section{Discussion}\label{sec:discussion}
We have established that the potential energy and the incoherent ISF show minor deviations from the triangular relation, while the MSD satisfies it within the numerical uncertainty. What constitutes a large enough deviation from the triangular relation to no longer allowing for the definition of a meaningful material time remains an open question. 

Our main finding is that the best material time to collapse the aging data is the property's ``own material'' time. The degree of improvement compared to the collapse obtained using the two other material times varies a lot, however. Thus while significant improvement in the collapse is obtained after reparametrizing the MSD by its own material time, illustrating the coherence between the existence of a material time and the degree to which the triangular relation applies, the collapses of $C_{uu}$ and of the incoherent ISF remain imperfect. 

In regard to the fact that the MSD-based material time does not collapse well the potential-energy data, this may reflect the previously reported fact that the slowest particles are most important for the overall structural relaxation \cite{dou22}; thus in future works it would be interesting to investigate whether the use of the ``harmonic inherent MSD'' of Ref. \onlinecite{dou22}, which emphasizes the slow particles, results in a better collapse of the current large up jump data.

Another point that deserves consideration in future works is whether the existence of dynamical heterogeneities imply that there is no ``global clock'' in the case of large temperature up jumps. The possible existence of local clocks was discussed by Castillo and Parsaeian some time ago \cite{Castillo2007, Castillo2003, cas01} from the broader scope of identifying a time-reparameterization invariant action. More simulation data are needed, however, to elucidate whether the introduction of local clocks can extend the range of validity of material-time-based descriptions of physical aging.

\section*{APPENDIX}

This Appendix gives technical details on the analysis.

\subsection{Triangular relation algorithm}\label{app:binning_algo}

\begin{figure}[ht]
    \centering
    \includegraphics[width=\linewidth]{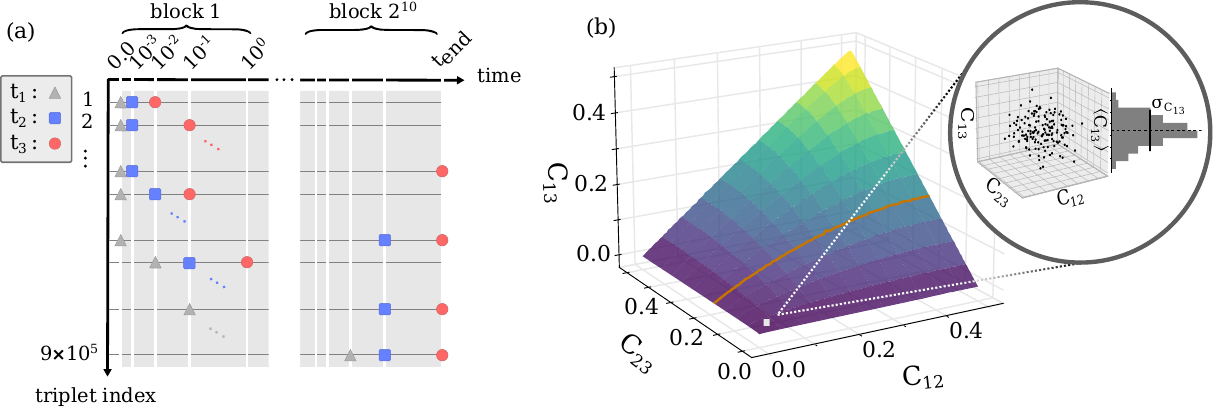}
    \caption{(a) Illustration of the sampling of time triplets used to calculate correlation ``triangles''. The values on the time axis illustrate the logarithmic increase in time intervals within a single simulation block.
    (b) Illustration of the triangle-binning algorithm. Each pixel contains a distribution of triangles $(C^\cA_{12}, C^\cA_{23}, C^\cA_{13})$. The orange line shows an example of the mean value of $C^\cA_{13}$ with respect to $C^\cA_{12}$ for a given value of $C^\cA_{23}$ as presented in 
    \fig{fig:triang_results}(a-c) (figure reproduced from Ref. \onlinecite{Amari2026}).}
    \label{fig:binning_algo}
\end{figure}

The triangular relation is investigated numerically by looking at many time-autocorrelation functions referring to pairs of times involving two of the three times $t_1 < t_2 < t_3$, measured after the thermostat temperature is changed at $t=0$. During aging, if the values of $C^\mathcal{A}_{12}$ and $C^\mathcal{A}_{23}$ are known, the triangular relation states that $C^\mathcal{A}_{13}$ is given. We selected 250 logarithmically spaced values of $t_1$ and 60 values of $\tau$ (logarithmically spaced for $C_{uu}$ and ISF, and linearly spaced for the MSD) in order to obtain a spectrum of $C^{\cA}_{12}$ and $C^{\cA}_{23}$ values that is as uniform as possible. Calling the set of $\tau$ values $M$, for each $t_1$ we identified $t_2 = t_1 + \tau'$ in which $\tau' \in M$, and for each $t_2$ we put $t_3 = t_2 + \tau''$ with $\tau'' \in M$. This gave a total of $250 \times 60^2 = 900 \,000$ $(t_1, t_2, t_3)$ triplets. We left out those for which $t_3$ is outside the total simulation time.  

Aging behavior appears at long times, while at short times the system shows vibrations inside cages formed by nearby particles. In the short-time regime, the fluctuation-dissipation theorem holds corresponding to the bath temperature. We do not show these early-time results because they are not important for this study 
(a material-time description does not make the early-time data collapse). To investigate the triangular relation we divided both $C^\mathcal{A}_{12}$ and $C^\mathcal{A}_{23}$ into equal bins of size 0.005, thus creating a grid of pixels. Each pair $(C^\mathcal{A}_{12}, C^\mathcal{A}_{23})$ falls into one pixel, and in this way each pixel contains a distribution of “triangles” with sides $(C^\mathcal{A}_{12}, C^\mathcal{A}_{23}, C^\mathcal{A}_{13})$. We ignored pixels with fewer than 10 triangles, leaving 2500-10000 pixels. For those pixels, we calculated the average and standard deviation of $C_{13}$. The averages are shown in the figures of the main paper. The quality of these results depends on how the times are sampled and how the correlation changes with $t_1$ and $t_2$. Ideally, one would like to sample times more finely where the autocorrelation changes rapidly. Due to limitations of storage and the block structure of GPU MD simulations, we used a mixed approach: time was divided into several linear blocks, and within each block data are saved at logarithmic intervals.

We use the standard deviation of $C^\cA_{13}$, $\sigma^2 \equiv \ea{(C^\cA_{13})^2} - \ea{C^\cA_{13}}^2$, as an indicator of how well \eq{eq:triang} is obeyed. We define $\sigma_{\text{uniform}}$ as the deviation only due to the variation of the mean value of $F^\cA$ in \eq{eq:triang} within the given pixel, while  $\sigma_{\text{true}}$ is the standard deviation of $C^\cA_{13}$ values within the pixel due to numerical noise and deviations from \eq{eq:triang}. The accessible quantity is $\sigma = \sigma_{\text{true}} + \sigma_{\text{uniform}}$; the ``small-pixel'' limit is reached when $\sigma_{\text{true}}>>\sigma_{\text{uniform}}$ for all pixels. The standard deviation of a continuous uniform distribution across a range $\Delta C^\cA_{13}$ is given by  \cite{Ross1998}
\be
    \sigma_{\text{uniform}} = \frac{\Delta C^\cA_{13}}{2\sqrt{3}} = \frac{\Delta C^\cA_{13}}{\Delta C^\cA_{12}} \frac{\Delta C^\cA_{12}}{2\sqrt{3}} 
\ee
where $\frac{\Delta C^\cA_{13}}{\Delta C^\cA_{12}}$ approximates $\frac{dC^\cA_{13}}{dC^\cA_{12}}$, the slope of $C^\cA_{13}[C^\cA_{12}]$. Thus the pixel size $\Delta C^\cA_{12}$ should satisfy
\be
    \Delta C^\cA_{12} << 2\sqrt{3} \; \sigma_{\text{true}}\; \Big( \frac{dC^\cA_{13}}{dC^\cA_{12}} \Big)^{-1}
\ee
From Ref. \onlinecite{Amari2026} for $\cA=u$ we find $\frac{dC^\cA_{13}}{dC^\cA_{12}}\leq 1$ and $\sigma_{\text{true}} < 10^{-2}$ so we need to satisfy $ \Delta C^\cA_{12} << 3\times 10^{-2}$. Based on this we took the pixel side length to be $\Delta C^\cA_{12} = 5\times 10^{-3}$.

\subsection{Triangular-relation heat maps}

\begin{figure}[ht!]
    \centering
    \includegraphics[width=1\linewidth]{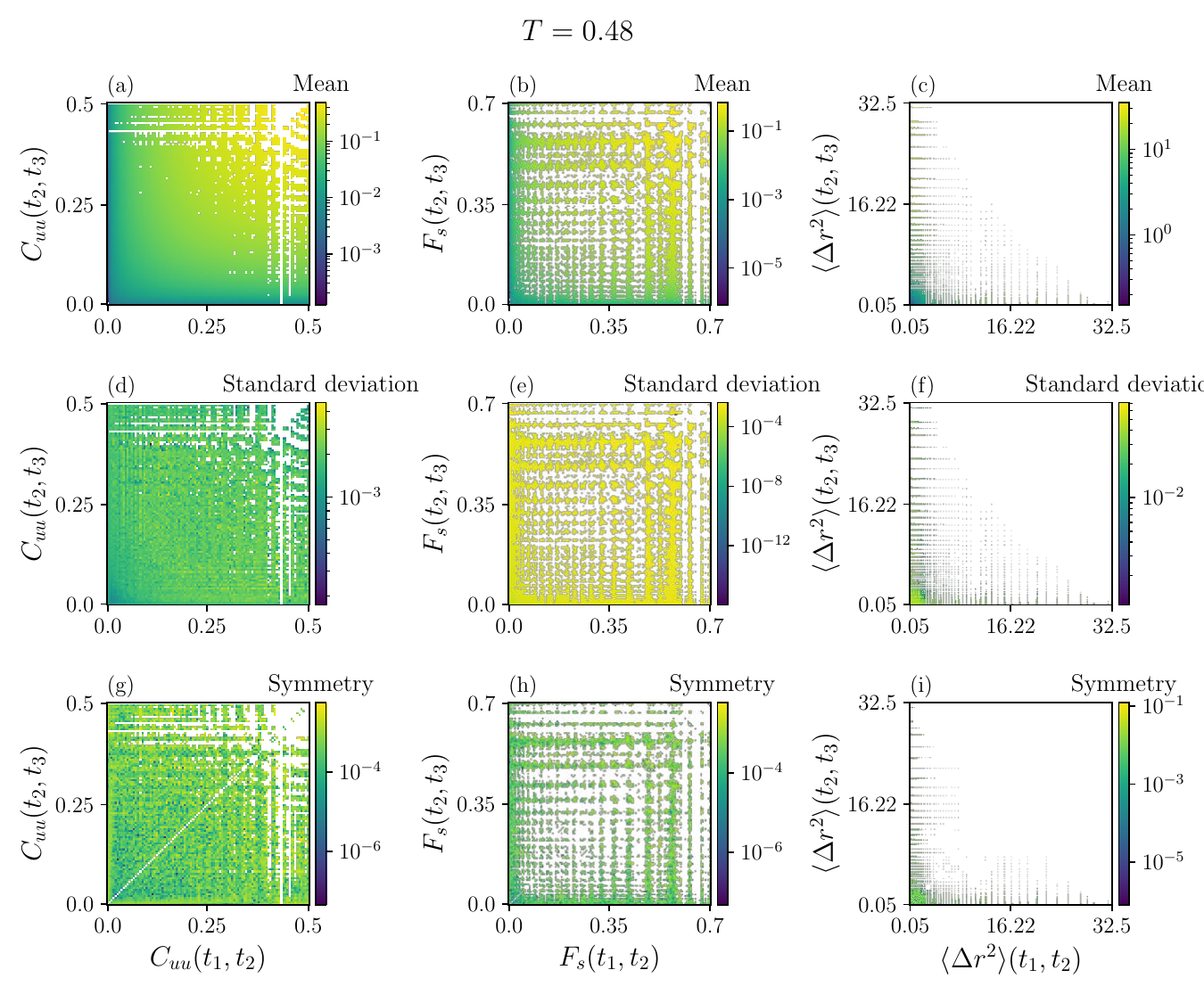}
    \caption{Parametric representation of the triangular relation at equilibrium for (left) the potential-energy time-autocorrelation function, (middle) the incoherent ISF, and (right) the MSD. (a-c) Per pixel average, (d-f) standard deviation, and (g-i) deviation from symmetry.}
    \label{fig:tri_heatmap_1}
\end{figure}

\begin{figure}[H]
    \centering
    \includegraphics[width=1\linewidth]{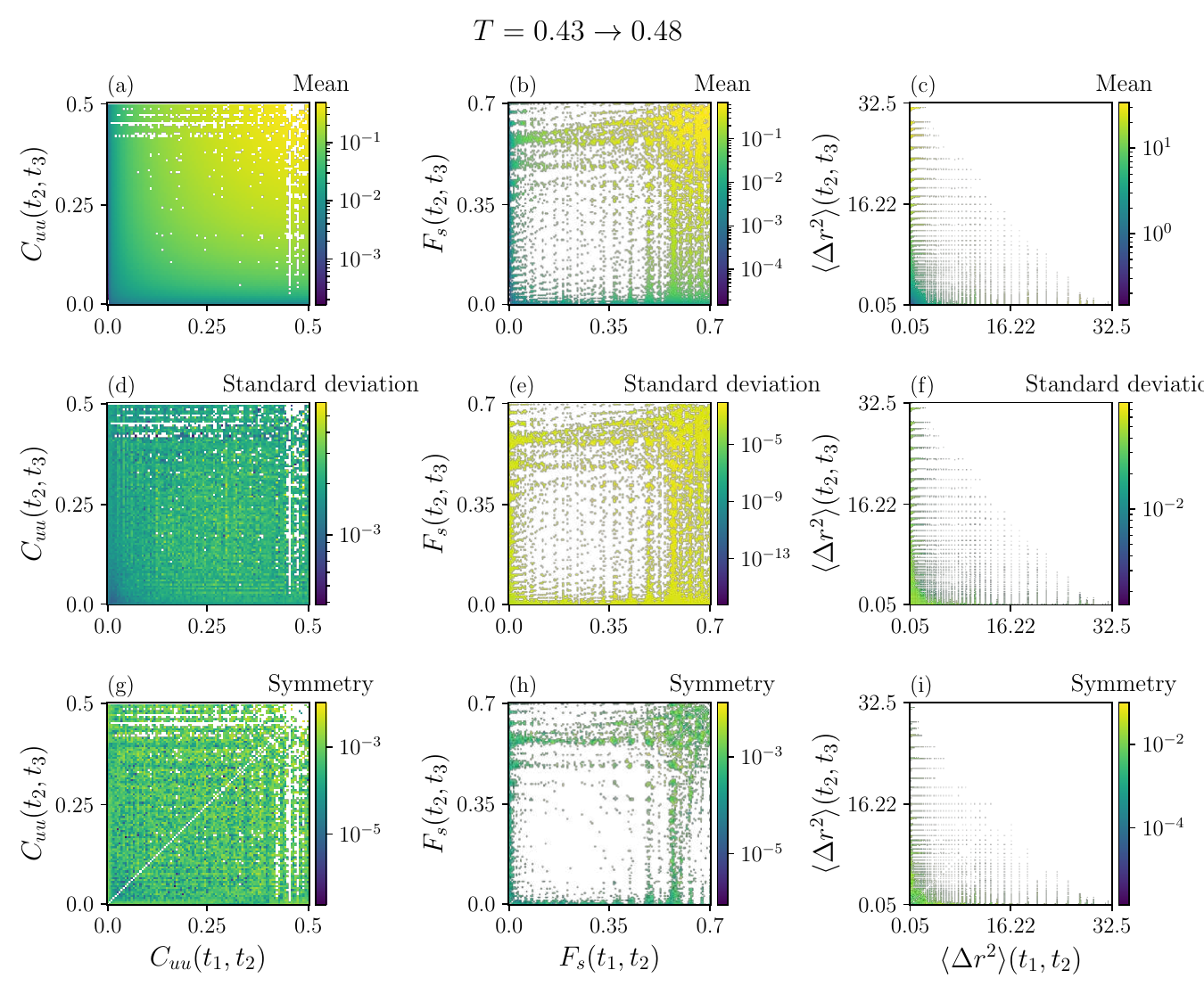}
    \caption{Triangular relation during aging after the jump from $T=0.43$ to $T=0.48$ for (left) the potential-energy time-autocorrelation function, (middle) the incoherent ISF, and (right) the MSD. (a-c) Per pixel average of, (d-f) standard deviation, and (g-i) deviation from symmetry.}
    \label{fig:tri_heatmap_2}
\end{figure}

\begin{figure}[H]
    \centering
    \includegraphics[width=1\linewidth]{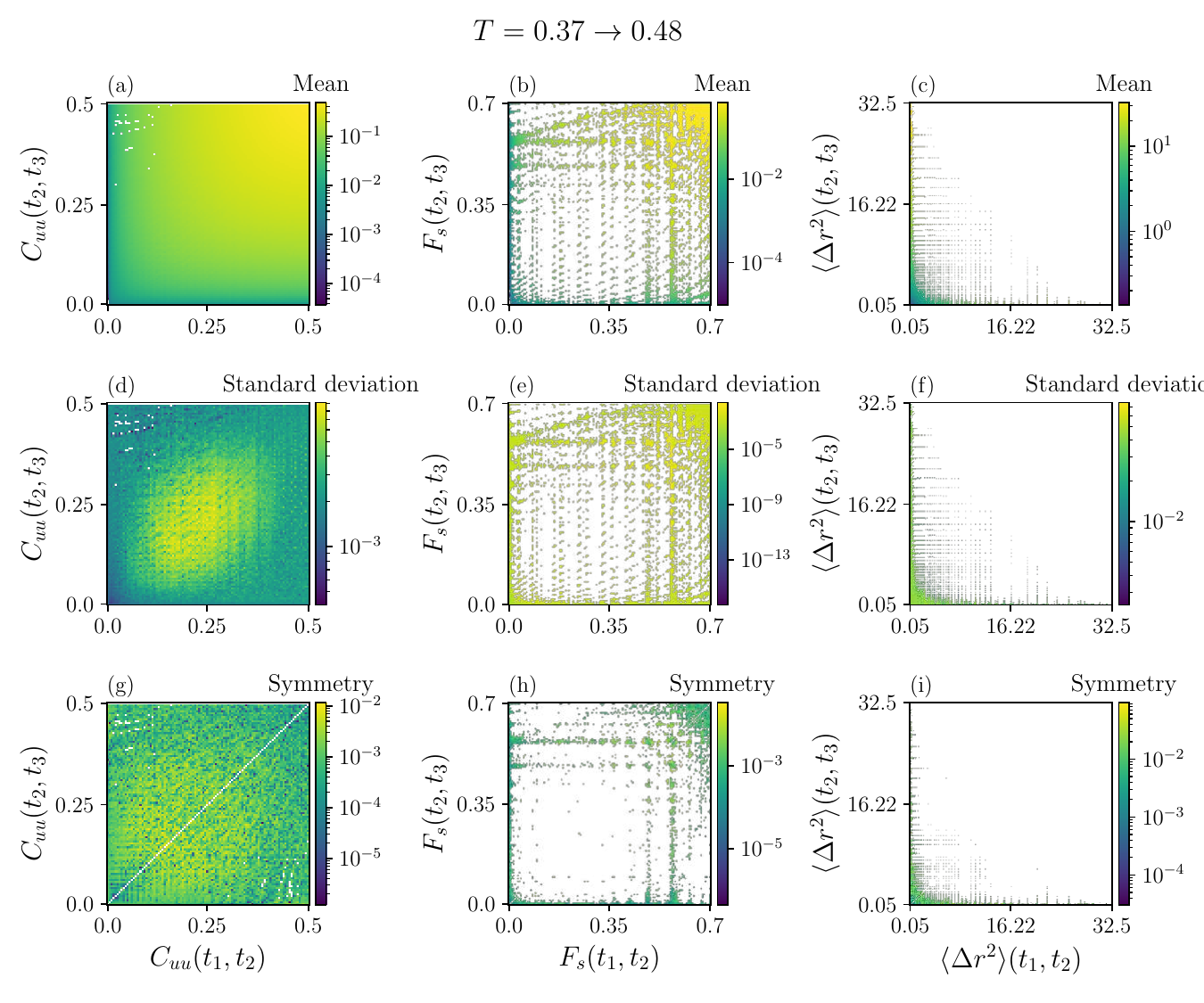}
    \caption{Triangular relation during aging after the jump from $T=0.37$ to $T=0.48$ for (left) the potential-energy time-autocorrelation function, (middle) the incoherent ISF, and (right) the MSD. (a-c) Per pixel average of, (d-f) standard deviation, and (g-i) deviation from symmetry.}
    \label{fig:tri_heatmap_3}
\end{figure}

\Crefrange{fig:tri_heatmap_1}{fig:tri_heatmap_3} show heat maps of the per pixel average, $\ea{C^{\cA}_{13}}_p$, standard deviation, $\sigma_{C^{\cA}_{13}}$, and deviation from symmetry, $|\ea{C^{\cA}_{13}}_p - \ea{C^{\cA}_{31}}_p|$, for all values of $C^\cA_{12}$ and $C^\cA_{23}$ for pixels containing at least 10 data points. For plotting and illustration purposes, the MSD heat maps are shown for pixel size 0.12 instead of the 0.005 used in the data analysis of the main text. White denotes disregarded pixels or (more rarely) the location where the plotted value is exactly zero.


\subsection{The quantity $a$ defining one unit of material time}\label{app:fixepoint_study}

\begin{figure}[ht]
    \centering
    \includegraphics[width=1\linewidth]{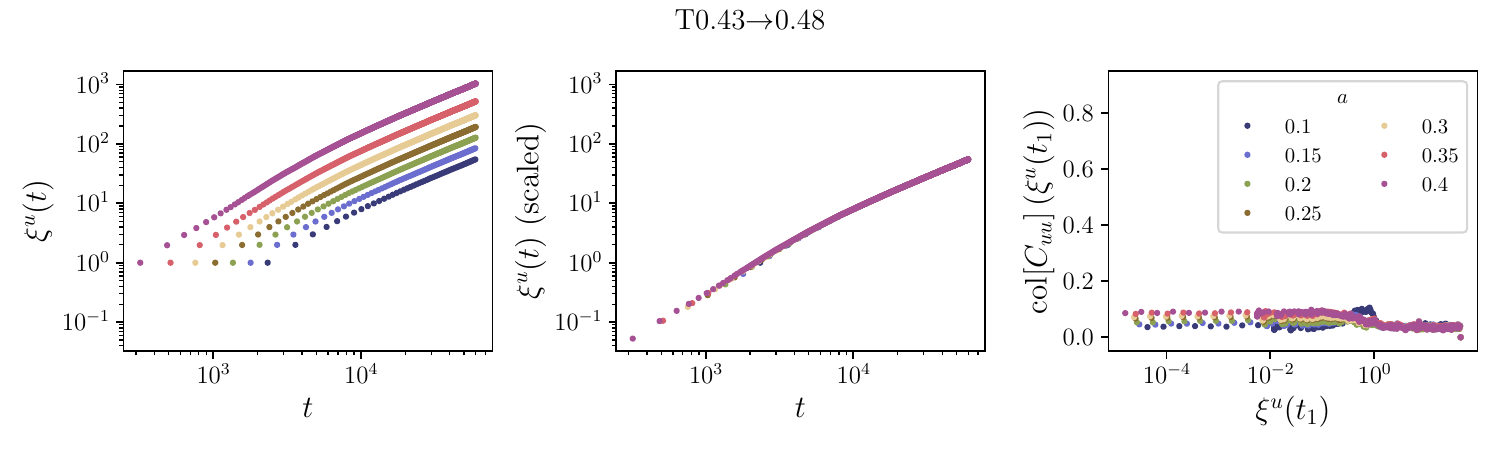}
    \includegraphics[width=1\linewidth]{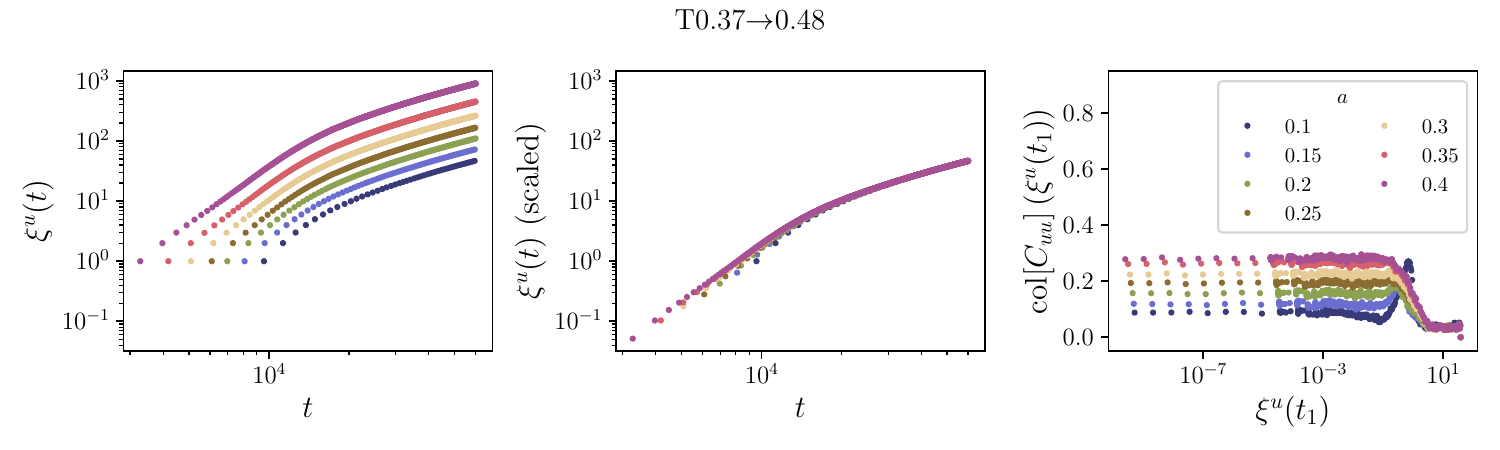}
    \caption{Material time defined from $C_{uu}$ for values of $a$ between 0.1 and 0.4. The behavior of the collapse (right panels) varies from monotonically decreasing for large values of $a$ to peaking at intermediate times for small $a$. Top row: moderate jump; Bottom row: large jump. 
    Left: Material time as a function of time; 
    Middle: Material time scaled to match at equilibrium at $T=0.48$; 
    Right: Quantification of the collapse of each curve $C_{uu}(\xi^u_2-\xi^u_1; \xi^u_1)$ with the final equilibrium $\phi^u_\text{eq}(t_2-t_1)$ for each value of $\xi^u_1$.}
    \label{fig:fixepoint_Cuu}
\end{figure}

\begin{figure}[ht]
    \centering
    \includegraphics[width=1\linewidth]{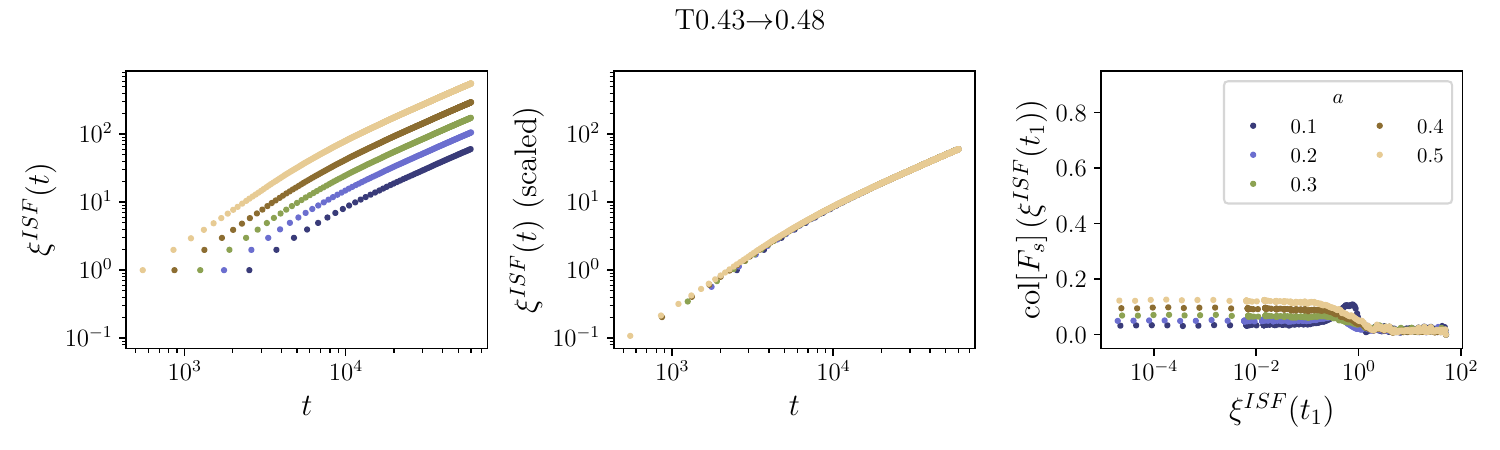}
    \includegraphics[width=1\linewidth]{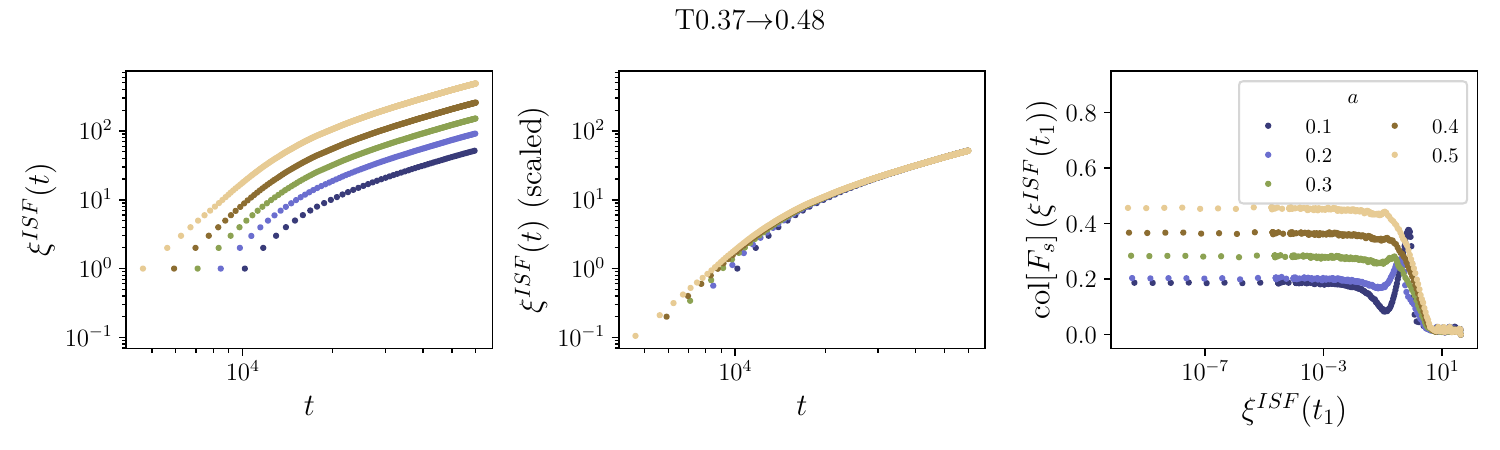}
    \caption{Material time defined from $F_{s}$ for values of $a$ between 0.1 and 0.5. The behavior of the collapse (right panels) varies from monotonically decreasing for large values of $a$ to peaking at intermediate times for small $a$. Top row: moderate jump; Bottom row: large jump. 
    Left: Material time as a function of time; 
    Middle: Material time scaled to match at equilibrium at $T=0.48$; 
    Right: Quantification of the collapse of each curve $F_{s}(\xi^{ISF}_2-\xi^{ISF}_1; \xi^{ISF}_1)$ with the final equilibrium $\phi^{ISF}_\text{eq}(t_2-t_1)$ for each value of $\xi^{ISF}_1$.}
    \label{fig:fixepoint_Fs}
\end{figure}

\begin{figure}[ht]
    \centering
    \includegraphics[width=1\linewidth]{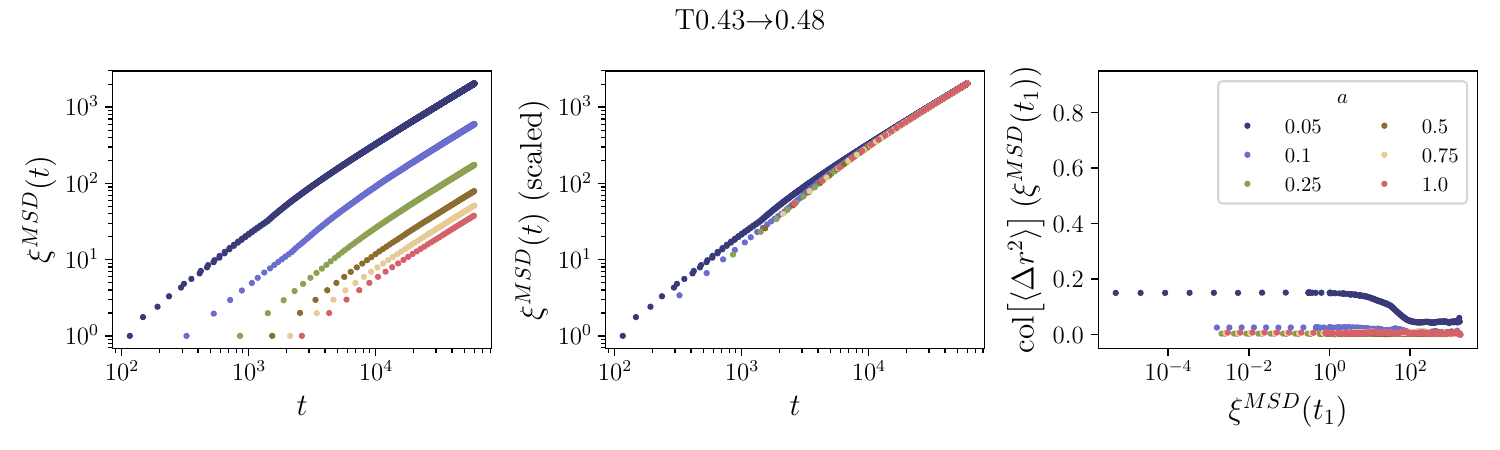}
    \includegraphics[width=1\linewidth]{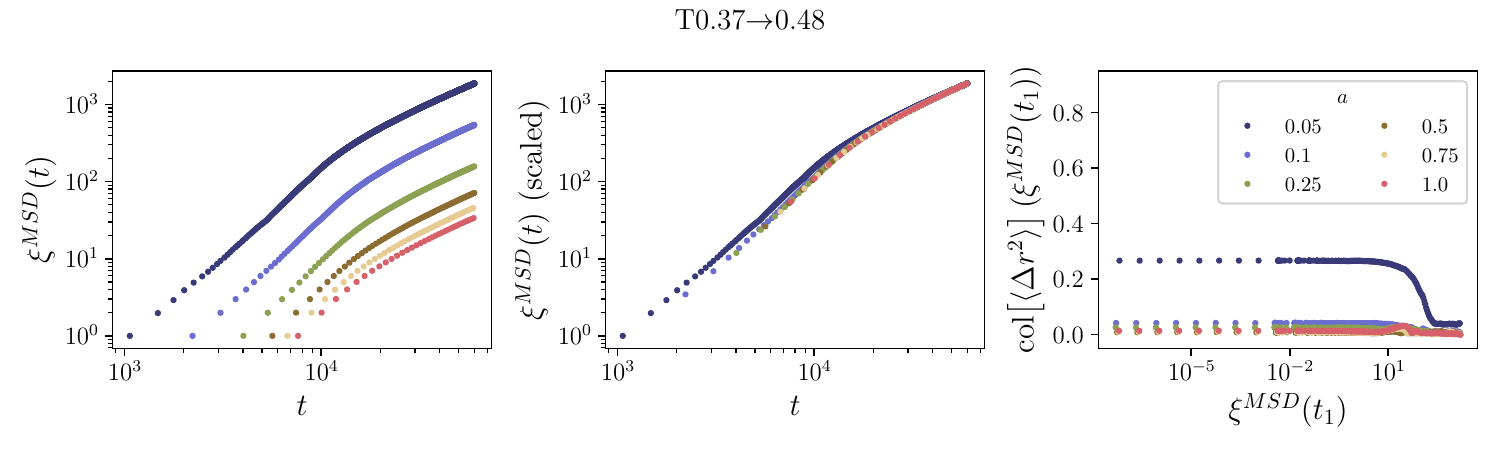}
    \caption{Material time defined from the MSD for values of $a$ between 0.1 and 0.5. The behavior of the collapse (right panels) varies from monotonically decreasing for large values of $a$ to peaking at intermediate times for small $a$. Top row: moderate jump; Bottom row: large jump. 
    Left: Material time as a function of time; 
    Middle: Material time scaled to match at equilibrium at $T=0.48$; 
    Right: Quantification of the collapse of each curve $\ea{\Delta r^2}(\xi^{MSD}_2-\xi^{MSD}_1; \xi^{MSD}_1)$ with the final equilibrium $\phi^{MSD}_\text{eq}(t_2-t_1)$ for each value of $\xi^{MSD}_1$.}
    \label{fig:fixepoint_msd}
\end{figure}

The choice of $a$, the amount of decorrelation defining one unit of material time, is in principle arbitrary. In practice, however, due to  deviations from the exact triangular relation as well as numerical uncertainties, different choices of $a$ result in slightly different material times. This, of course, affects the degree of collapse of the aging response after reparameterization of time. The behavior of the normalized Euclidean norm of the deviation from the final equilibrium of all aging curves, \eq{eq:def_L2norm}, changes from monotonically decreasing (for values of $a$ close to the plateau) to an almost oscillating collapse (as shown in \crefrange{fig:fixepoint_Cuu}{fig:fixepoint_msd} for $F_s$ for the jump $T=0.37 \rightarrow 0.48$ for $a=0.1$ where $col[F_s]$ starts at a low value, then goes down before increasing to reach a maximum at intermediate values of $\xi(t_1)$ and finally decreases to a minimum at long times.  

\bibliography{aude_ref}
\end{document}